# Analytical Modeling of Participation Reduction in Superconducting Coplanar Resonator and Qubit Designs through Substrate Trenching

Conal E. Murray[*]

*Abstract*— A strategy aimed at decreasing dielectric loss in coplanar waveguides (CPW) and qubits involves the creation of trenches in the underlying substrate within the gaps of the overlying metallization. Participation of contamination layers residing on surfaces and interfaces in these designs can be reduced due to the change in the effective dielectric properties between the groundplane and conductor metallization. Although finite element method approaches have been previously applied to quantify this decrease, an analytical method is presented that can uniquely address geometries possessing small to intermediate substrate trench depths. Conformal mapping techniques produce transformed CPW and qubit geometries without substrate trenching but a non-uniform contamination layer thickness. By parametrizing this variation, one can calculate surface participation through the use of a two-dimensional, analytical approximation that properly captures singularities in the electric field intensity near the metallization corners and edges. Examples demonstrate two regimes with respect to substrate trench depth that capture an initial increase in substrate-to-air surface participation due to the trench sidewalls and an overall decrease in surface participation due to the reduction in the effective dielectric constant, and are compared to experimental measurements to extract loss tangents on this surface.

## TABLE I
### PARAMETER VALUES

| | |
|---|---|
| $a$ | half of the centerline conductor width (CPW) |
| | half of the capacitor gap (qubit) |
| $b$ | half of the distance between groundplane metallization (CPW) |
| | half of the distance between outer capacitor edges (qubit) |
| $\delta_i$ | contamination layer thickness |
| $\varepsilon_i$ | relative dielectric constant of material i |
| $K(k)$ | complete elliptic integral of the first kind with modulus k |
| $K'(k)$ | complement of $K(k)$ |
| $k = a / b$ | modulus associated with $K$ and $K'$ |
| $P_i$ | surface participation associated with interface i |
| $\theta_a, \theta_b$ | integrands associated with surface participation calculations |
| $t$ | trench depth |
| $u'_i$ | location of metallization edges in the transformed geometry |
| $u_i$ | location of trench bottom corners in the transformed geometry |
| $w = \xi + i\, \eta$ | coordinate system of the trenched geometry |
| $z = x + i\, y$ | coordinate system of the transformed geometry |

## I. INTRODUCTION

Coplanar waveguide (CPW) resonators have been an integral component of signal transmission for over half of a century, spanning a wide range of technologies from monolithic microwave integrated circuitry (MMIC) [1] to superconducting applications such as kinetic inductance detectors [2], [3] and quantum computing [4], [5]. Assessing various mechanisms of loss within microwave resonator designs is critical to ensuring high quality factors and increased efficiency in such devices. Dielectric loss is known to impact the quality factors of both CPW and qubit designs, where the electric fields emanating from metallization features penetrate the surrounding materials and interact with two level systems [6], [7]. Such materials include the substrate as well as contamination layers (e.g., oxides) that exist or may form along key surfaces and interfaces. Methods to reduce this loss can involve improvements in the treatment of surface and interfaces [8]–[10], as well as the incorporation of materials with fewer species of impurities known to be detrimental to resonator quality factors [7], [11], [12]. However, the intrinsic effect that drives surface participation: the electric field energy, can also be reduced by adjusting the design geometry [8], [13], [14]. For example, the creation of trenches within silicon substrates possessing overlying CPW metallization has been accomplished by etching recesses in the gaps between the centerline conductor and groundplanes. This technique has been demonstrated to reduce RF loss and leakage current in Si-based MMIC CPW's [15] in addition to increasing the quality factor in several examples associated with quantum computing [9], [16]-[20].

Previous treatments to calculate the effects of substrate trenching often involve finite-element-method (FEM) based models to simulate electric field energy along specific surfaces of resonator designs [17], [19]-[21]. While FEM approaches are versatile with respect to the variety of geometries they can analyze, their use in predicting surface participation often requires power law approximations of the electric field distributions that possess singularities at the metallization edges or the scaling of contamination layer thicknesses [20] from solutions at larger dimensions. For small trench depths, a logarithmic dependence of surface participation with substrate trench depth is predicted [14] due to the square root dependence

 

of electric field intensity with distance from the edge of a metallization sheet [22]. However, in the limit of zero trench depth, the surface participation must converge to a finite value, as demonstrated through an analytical approximation based on the conformal mapping of untrenched coplanar designs [23]. In Sections II and III, we present analytical, closed-form solutions to calculate surface participation for CPW and coplanar capacitors, respectively, that provide a necessary link between untrenched substrate geometries and trenched designs, validated by FEM simulations at larger dimensions. This formalism provides a unique way of quantifying surface participation in shallow trenches, applicable to many resonator and qubit builds.

## II. COPLANAR WAVEGUIDES

### A. Model

We begin with a quasi-static treatment of the electric field distributions generated in an arbitrary, two-dimensional cross-section through a CPW geometry. The metallization is assumed to behave as a perfect electric conductor (PEC) so that the electric fields are oriented normal to the conductor surfaces and magnetic fields do not penetrate the metallization. Fig. 1(a) depicts a cross-sectional schematic of this geometry in which the gap widths between the centerline conductor of width $2a$ and the groundplanes are equal to $b - a$ in the complex w-plane ($w = \xi + i\,\eta$). Due to the intrinsic symmetry of the design about the line $\xi = 0$ that bisects the centerline conductor, only half of the geometry is shown ($\xi > 0$) in detail. In the gap between the PEC metallization, the underlying substrate, with relative dielectric constant $\varepsilon_{sub}$, is trenched to a depth $\eta = t$. Within this geometry, contamination layers of constant thickness, $\delta_0$, and relative dielectric constant, $\varepsilon_c$, can be located on the substrate-to-metal (SM) interface, the substrate-to-air (SA) interface or the metal-to-air (MA) interface. Conformal mapping, where

$$w = \int \sqrt{\frac{(z+u_1)(z-u_1)(z+u_2)(z-u_2)}{(z+u'_1)(z-u'_1)(z+u'_2)(z-u'_2)}}\,dz \tag{1}$$

maps the trenched geometry possessing right angles in the $w$ plane to a half-space within the complex $z$ plane ($z = x + i\,y$) using the Schwarz-Christoffel transformation. This flattening of the CPW design, as shown in Fig. 1(b), results in an effective reduction in the centerline conductor width, defined as $2u'_1$, and an increase in the metallization gap, $u'_2 - u'_1$. The values corresponding to the corner positions ($u_i$ and $u'_i$) in the transformed half-space are not known *a priori* and must be calculated through iteration by integrating (1) over a known trench depth and metallization gap $b - a$ to converge to the correct values. An approximate formulation can also be generated for shallow trenches ($t \ll a$):

$$u'_1 \sim a - \frac{t}{\pi}\left[1 + \ln(4\pi) + \ln\left(\frac{1-k}{1+k}\right) - \ln\left(\frac{t}{a}\right)\right] \tag{2}$$

$$u'_2 \sim b + \frac{t}{\pi}\left[1 + \ln(4\pi) + \ln\left(\frac{1-k}{1+k}\right) - \ln\left(\frac{t}{b}\right)\right] \tag{3}$$

where $u_1$ and $u_2$ can be represented by $u_1 \sim u'_1 + 2t/\pi$, $u_2 \sim u'_2 - 2t/\pi$, as shown in Appendix A.

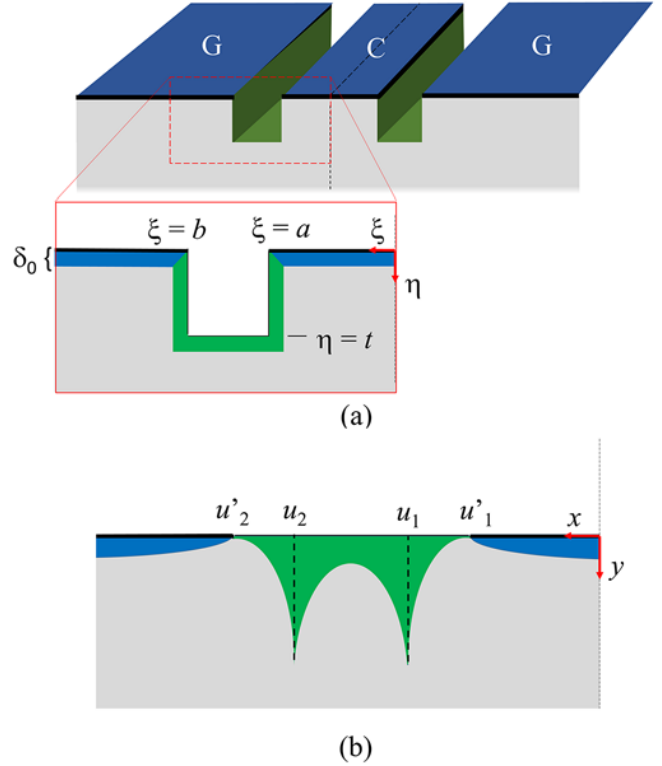

FIG. 1 (a) Cross-sectional geometry of a coplanar waveguide (CPW) resonator possessing a trenched substrate between the centerline conductor ('C') and groundplane metallization ('G'). In the inset to Fig. 1(a), the solid region represents a hypothetical contamination layer of constant thickness, $\delta_0$, where the blue regions correspond to the substrate-to-metal (SM) interfaces and the green region the substrate-to-air (SA) interfaces. (b) Transformed CPW geometry of the Fig. 1(a) inset, which removes trenching but possesses a variable contamination layer thickness, $\delta_t(x)$, due to conformal mapping (1).

Note that the contamination layer thickness in the transformed design (Fig. 1(b)) significantly decreases near the edges of the transformed metallization ($u'_1$, $u'_2$) and increases near the bottom corners of the trench ($u_1$, $u_2$). In the region near $u'_1$, we can approximate the conformal map:

$$w = \int \sqrt{\frac{(z+u_1)(z-u_1)(z+u_2)(z-u_2)}{(z+u'_1)(z-u'_1)(z+u'_2)(z-u'_2)}}\,dz$$

$$\sim \sqrt{\frac{(x+u_1)(x-u_1)(x+u_2)(x-u_2)}{(x+u'_1)(x+u'_2)(x-u'_2)}}\int\frac{dz}{\sqrt{x+iy-u'_1}} \tag{4}$$

and integrate over the path corresponding to the contamination layer thickness of the original design, $\delta_0$ (Fig. 1(a)) to obtain (see Appendix B):

$$\delta_1(x) \sim$$
$$\frac{\delta_0}{2}\left[\frac{(u'^2_2-x^2)(u'_1+x)}{(u^2_2-x^2)\,|\,u^2_1-x^2\,|}\right]\sqrt{\delta_0^2 + 4\frac{(u^2_2-x^2)\,|(u^2_1-x^2)(u'_1-x)|}{(u'^2_2-x^2)(u'_1+x)}} \tag{5}$$

A similar approximation of the conformal mapping near $u'_2$ yields:

$$\delta_2(x) \sim$$

$$\frac{\delta_0}{2} \left[ \frac{(x^2 - u_1'^2)(x + u_2')}{(x^2 - u_1^2) \, |x^2 - u_2^2|} \right] \sqrt{\delta_0^2 + 4 \frac{(x^2 - u_1^2)[(x^2 - u_2^2)(u_2' - x)]}{(x^2 - u_1'^2)(x + u_2')}} \quad (6)$$

The contamination layer thickness along the SM interface can be represented by $\delta_1(x)$ underneath the centerline conductor ($0 \leq |x| \leq u'_1$) and $\delta_2(x)$ under the groundplane ($|x| \geq u'_2$). For the SA interface, (5) holds as an approximation along the inner trench sidewall and (6) along the outer trench sidewall. The trench bottom can be separated into two regions, $u_1$ to $(u_1 + u_2)/2$ and $(u_1 + u_2)/2$ to $u_2$, in which (5) and (6) are employed, respectively.

For a CPW without substrate trenching, the surface participation along key interfaces can be approximated as [23]:

$$P_{SA} \sim \frac{\varepsilon_{c:SA}}{(\varepsilon_{sub} + 1)} \frac{1}{2(1 - k)K'(k)K(k)}$$
$$\cdot \left( \frac{\delta_0}{a} \right) \left\{ \ln \left[ 4 \left( \frac{1 - k}{1 + k} \right) \right] - \frac{k \ln(k)}{(1 + k)} + 1 - \ln \left( \frac{\delta_0}{a} \right) \right\} \quad (7)$$

$$P_M \sim \frac{\varepsilon_{sub}^2}{\varepsilon_{c:SM} \varepsilon_{c:SA}} P_{SA} \quad (8)$$

$$P_{MA} \sim \frac{\varepsilon_{c:SM}}{\varepsilon_{c:MA}(\varepsilon_{sub}^2)} P_M \quad (9)$$

where $K(k)$ and $K'(k)$ represent the complete elliptic integral of the first kind and its complement, respectively, the modulus $k = a/b$, and $\varepsilon_{c:SM}$, $\varepsilon_{c:SA}$ and $\varepsilon_{c:MA}$ are the relative dielectric constants associated with contamination layers present at the SM, SA and MA interfaces, respectively. Equations (7) to (9) were derived by first integrating the electric field energy as a function of distance across the resonator structure at a finite depth, $y$, then integrating with respect to $y$ from the top surface of the substrate ($y = 0$) to a depth corresponding to the contamination layer thickness, $\delta_0$. We can apply a similar methodology to the transformed geometry depicted in Fig. 1(b), where the order of integration must be switched due to the variable contamination layer thickness, $\delta_i(x)$, as parametrized in (5) and (6). The resulting equations for estimating surface participation in trenched substrates take the form:

$$P_{SA}^t \sim \frac{\varepsilon_{c:SA}}{(\varepsilon_{sub} + 1)} \frac{u_2'^2}{K \left( \frac{u_1'}{u_2} \right) K' \left( \frac{u_1'}{u_2} \right)}$$
$$\cdot \left[ \int_{u_1'}^{\frac{(u_1 + u_2)}{2}} \theta_a(x) \, dx + \int_{\frac{(u_1 + u_2)}{2}}^{u_2} \theta_b(x) \, dx \right] \quad (10)$$

$$P_{SM}^t \sim \frac{\varepsilon_{sub}^2}{\varepsilon_{c:SM}(\varepsilon_{sub} + 1)} \frac{u_2'^2}{K \left( \frac{u_1'}{u_2} \right) K' \left( \frac{u_1'}{u_2} \right)}$$
$$\cdot \left[ \int_0^{u_1'} \theta_a(x) \, dx + \int_{u_2}^\infty \theta_b(x) \, dx \right] \quad (11)$$

$$P_{MA}^t \sim \frac{1}{\varepsilon_{c:MA}(\varepsilon_{sub} + 1)} \frac{u_2'^2}{K \left( \frac{u_1'}{u_2} \right) K' \left( \frac{u_1'}{u_2} \right)}$$
$$\cdot \left[ \int_0^{u_1'} \theta_a(x) \, dx + \int_{u_2}^\infty \theta_b(x) \, dx \right] \quad (12)$$

where

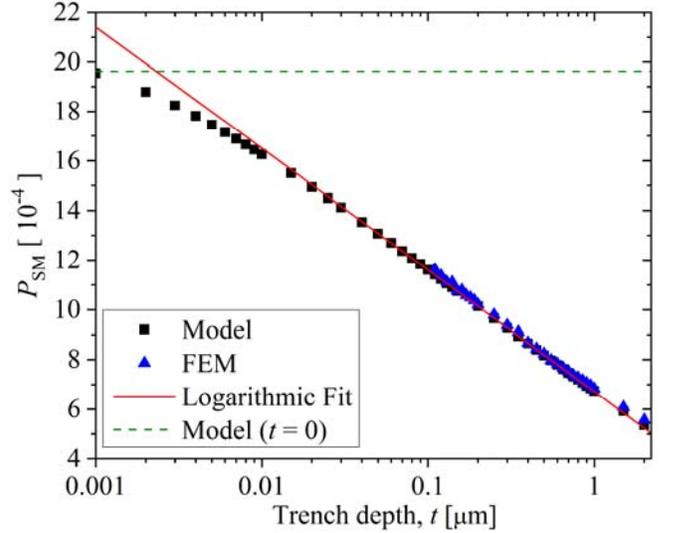

Fig. 2. Simulated SM surface participation for a CPW resonator design with a centerline conductor width of 10 μm ($a$ = 5 μm) and gap of 6 μm ($b$ = 11 μm), assuming a 2 nm thick contamination layer with a relative dielectric constant of 5.0. Squares correspond to the analytical model (11) and triangles those based on FEM modelling. The fit to the FEM results (solid line) illustrates the logarithmic dependence of both approaches which deviate at very shallow trench depths. In the limit of zero trench depth, the analytical model converges to the surface participation value calculated for a CPW resonator on a substrate without trenches (dotted line).

$$\theta_a(x) = \frac{1}{(u_2'^2 - x^2)(x + u_1')}$$
$$\cdot \left\{ \log[\delta_1(x) + \sqrt{(u_1' - x)^2 + [\delta_1(x)]^2}] - \log|u_1' - x| \right\} \quad (13)$$

$$\theta_b(x) = \frac{1}{(x^2 - u_1'^2)(x + u_2')}$$
$$\cdot \left\{ \log[\delta_2(x) + \sqrt{(x - u_2')^2 + [\delta_2(x)]^2}] - \log|x - u_2'| \right\} \quad (14)$$

### B. Results

Fig. 2 contains the calculated SM surface participation of a 2 nm thick contamination layer for a CPW with a 10 μm wide centerline conductor ($a$ = 5 μm) and gap of 6 μm ($b$ = 11 μm) as a function of trench depth, $t$. The relative dielectric constant of the substrate is assumed to be 11.45 and that of all the contamination layers are 5.0. The squares represent the values calculated by (11) while the triangles illustrate those simulated by FEM using Ansys HFSS (Ansys, Canonsburg, PA). Both approaches capture the pivotal trend that increasing trench depth, $t$, reduces surface participation and they yield good agreement at large trench depths. Unfortunately, it becomes more difficult for FEM modeling to capture the effects of singularities in the electric field intensity near the corners of the metallization for trenches shallower than 0.2 μm [23]. One can extrapolate these results by assuming a logarithmic dependence of surface participation with $t$ [14], as represented by the solid line in Fig. 2, illustrating the difference between the analytical model and logarithmic fit at trench depths below 10 nm. However, surface participation must be finite in the limit of zero trench depth, where the analytical approach converges to the untrenched value (0.00196) as derived using (7) and (8).

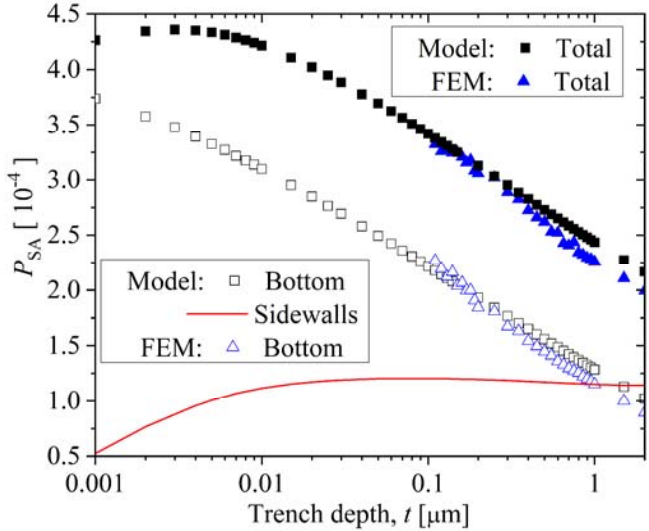

Fig. 3. Simulated SA surface participation for a CPW resonator design with a centerline conductor width of 10 μm ($a$ = 5 μm) and gap of 6 μm ($b$ = 11 μm), assuming a 2 nm thick contamination layer with a relative dielectric constant of 5.0. Squares correspond to the analytical model (10) and triangles to those based on FEM modelling. The difference (solid line) between the trench bottom and total SA values corresponds to the trench sidewall surface participation.

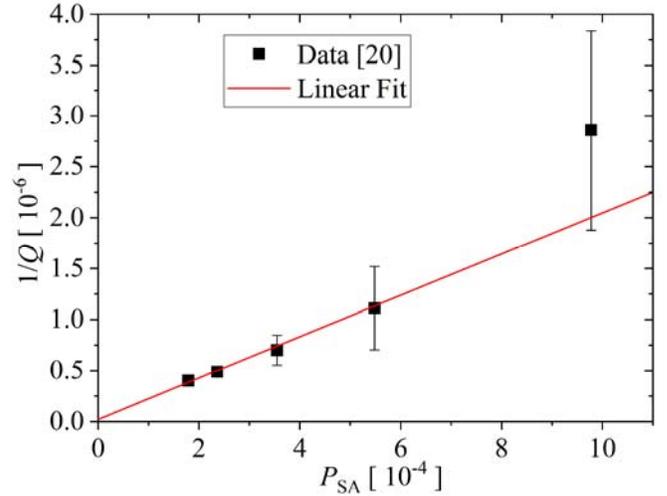

Fig. 4. Comparison of experimental, reciprocal Q values from [20] extracted from measurements conducted on CPW resonators on Si substrates with 150 nm deep trenches but different widths and calculated SA surface participation values using (10) assuming a 2 nm thick contamination layer with a relative dielectric constant of 5.0. Linear, least-squares fitting produces a loss tangent of $2.03 \pm 0.34$ x $10^{-3}$.

An analysis of SA surface participation for the same CPW design (10) reveals two regimes with respect to trench depth. As shown in the solid symbols of Fig. 3, surface participation increases from the untrenched value ($3.74 \times 10^{-4}$) for small values of $t$, followed by a decrease in a manner similar to SM surface participation. By separating the contribution solely from the trench bottom (open symbols), we find that the ratio of SM participation to the contribution of the SA trench bottom (5.24) is independent of trench depth for intermediate values of $t$ and equivalent to the ratio for untrenched designs, $\varepsilon^2_{SUB} / (\varepsilon_{C:SM} \varepsilon_{C:SA})$, predicted by (8). The increase in SA participation corresponds to the introduction of trench sidewalls, whose contribution is illustrated by the solid line in Fig. 3, that saturates at approximately 50 nm. Its influence is localized to a narrow region in trench depth, a crucial finding for quantifying dielectric loss in designs with extremely thin contamination layers (on the order of nanometers) and shallow substrate trenches.

We can demonstrate the utility of this model by applying its results to measurements conducted on CPW resonators. Fig. 4 contains a comparison of experimental, reciprocal $Q$ values [20] extracted from resonators on Si substrates possessing identical a / b ratios (1/2) and trench depths (150 nm) but a variety of conductor widths, 2a, ranging from 1.5 to 11 μm. These values represent the difference between experiments conducted at low photon number and high photon number, in which two-level systems saturate, to isolate dielectric loss from other potential loss mechanisms [20], and are plotted as a function of calculated SA surface participation (10) for a hypothetical, 2 nm thick contamination layer with a relative dielectric constant of 5.0. If we assume that the SA interface represents a dominant loss mechanism in the CPW resonators [24], [25], then $Q$ takes the form: $1/Q = 1/Q_0 + P_{SA} \tan(\delta_{SA})$, where $\tan(\delta_{SA})$ refers to the loss tangent of the SA contamination and $Q_0$ to loss effects independent of surface participation. This assumption is partly based on (8) and (9), indicating that the MA surface participation is substantially smaller than those of the SA or SM surfaces [23], and the finding that the calculated substrate participation (not shown) is anticorrelated with the experimental $Q$ values from [20]. Because loss due to SM surface participation cannot be ruled out, an estimate of the SA loss tangent based on the data in Fig. 4 represents an upper bound. Linear, least-squares fitting produces a loss tangent of $2.03 \pm 0.34$ x $10^{-3}$, and a $1/Q_0$ of zero within the fitting error. This $\tan(\delta_{SA})$ value is similar to that previously reported for contamination on silicon substrates [26], resides between those determined for thermally grown $SiO_2$ (3 x $10^{-4}$) and chemical vapor deposited amorphous $SiO_2$ (3x $10^{-3}$) [11], and below the upper bound of [21].

## III. COPLANAR CAPACITORS

### A. Model

The analytical approach presented in Section II can also be extended to coplanar capacitor designs, such as those incorporated in transmon qubits [14]. Fig. 5(a) illustrates a portion of such a structure, labelled 'C', where we assume a symmetric geometry exists about the $\xi = 0$ axis, and the groundplane is infinitely far away. The inset figure depicts a recessed substrate by a depth, $t$, from the capacitor metallization. Again, conformal mapping may be used to flatten this geometry to an untrenched half-space through the following transformation:

$$w = i \cdot t - \int \sqrt{\frac{(z+u_1)(z-u_1)(z+u_2)(z-u_2)}{(z+w_1)(z-w_1)(z+w_2)(z-w_2)}} \, dz \qquad (15)$$

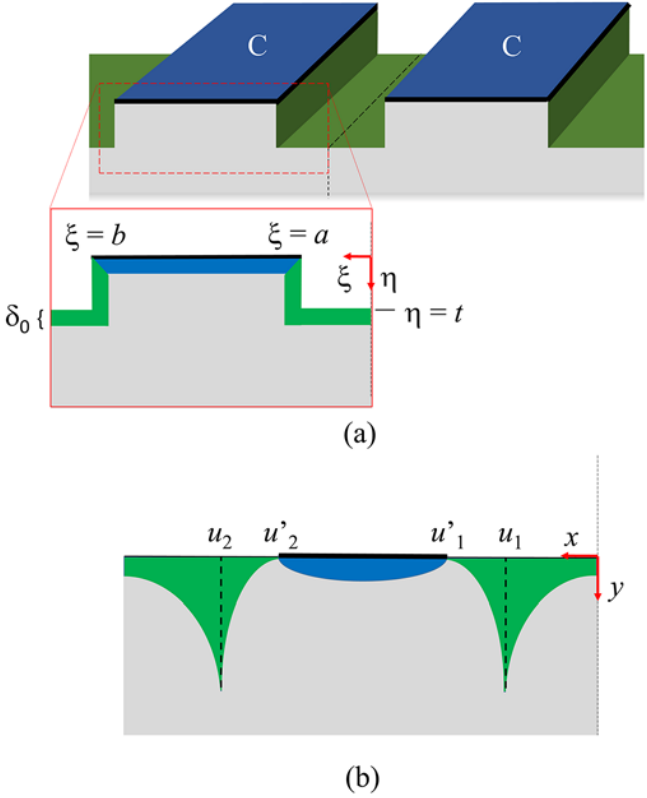

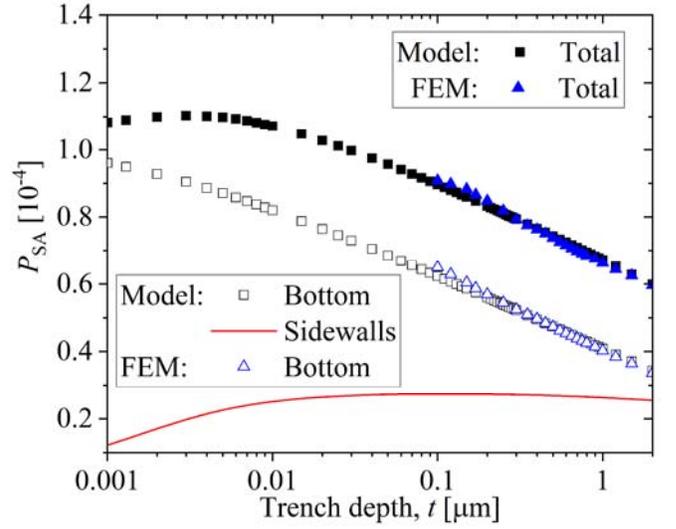

Fig. 6. Simulated SA surface participation as a function of trench depth for a transmon qubit ($a = 10$ μm, $b = 70$ μm) assuming a 2 nm thick contamination layer with relative dielectric constant of 5.0. Squares correspond to the numeric model (16) and triangles those based on FEM modelling. The difference (solid line) between the trench bottom and total SA values corresponds to the trench sidewall surface participation.

Fig. 5. (a) Cross-sectional schematic geometry of a section of coplanar capacitor ('C') design associated with transmon qubits where the substrate is trenched to a depth $\eta = t$. The blue region corresponds to the SM contamination layer and green region the SA contamination layer, both with thickness $\delta_0$. (b) Transformed qubit geometry of the Fig. 5(a) inset to that without a trench, where the contamination layer thickness $\delta_i(x)$ varies with position.

where $i \cdot t$ represents a constant placing the metallization surfaces at $\eta = 0$. One can readily recognize the similarity between (15) and the trenched CPW transformation (1) but now the contamination layer thickness in Fig. 5(b) exhibits maxima corresponding to the trench bottom corners ($u_i$) and minima at the metallization edges ($u'_i$). Therefore, an equivalent procedure in extracting surface participation values results in the following formulas:

$$P_{SA}^t \sim \frac{\varepsilon_{cSA}}{(\varepsilon_{sub}+1)} \frac{u'^2_2}{K\left(\frac{u'_1}{u'_2}\right)K'\left(\frac{u'_1}{u'_2}\right)} \left[\int_0^{u'_1} \theta_a(x)\,dx + \int_{u'_2}^{\infty} \theta_b(x)\,dx\right] \quad (16)$$

$$P_{SM}^t \sim \frac{\varepsilon_{sub}^2}{\varepsilon_{cSM}(\varepsilon_{sub}+1)} \frac{u'^2_2}{K\left(\frac{u'_1}{u'_2}\right)K'\left(\frac{u'_1}{u'_2}\right)}$$
$$\cdot \left[\int_{u'_1}^{\frac{(u'_1+u'_2)}{2}} \theta_a(x)\,dx + \int_{\frac{(u'_1+u'_2)}{2}}^{\infty} \theta_b(x)\,dx\right] \quad (17)$$

$$P_{MA}^t \sim \frac{1}{\varepsilon_{cMA}(\varepsilon_{sub}+1)} \frac{u'^2_2}{K\left(\frac{u'_1}{u'_2}\right)K'\left(\frac{u'_1}{u'_2}\right)}$$
$$\cdot \left[\int_{u'_1}^{\frac{(u'_1+u'_2)}{2}} \theta_a(x)\,dx + \int_{\frac{(u'_1+u'_2)}{2}}^{u'_2} \theta_b(x)\,dx\right] \quad (18)$$

where $\theta_a$ and $\theta_b$ are identical to those in (13) and (14).

## B. Results

Fig. 6 contains a comparison of SA surface participation values for a trenched qubit design similar to that corresponding to Mod D in [14] ($a = 10$ μm, $b = 70$ μm) to FEM simulations. Again, the relative dielectric constant of the substrate is assumed to be 11.45 and the contamination layers 5.0. The total SA surface participation (solid squares) follows the same trend as that observed in CPW designs, increasing from the untrenched value ($9.6 \times 10^{-5}$) as $t$ increases until approximately 50 nm where the contribution of the sidewalls (solid line) saturates. At larger trench depths, a logarithmic dependence is observed in both the trench bottom fraction (open squares) and the total SA surface participation, consistent with FEM modeling [14].

## IV. CONCLUSION

Analytical modeling of surface participation in trenched, coplanar structures provides a critical link between values associated with untrenched CPW or qubit designs and the logarithmic dependence with respect to trench depth exhibited by FEM-based analyses. In particular, this approach provides accurate values of surface participation for designs with contamination layer thicknesses and trench depths that extend to the nanometer range, which is directly relevant to current technology. The results, difficult to be assessed using FEM-based modeling, reveal new insights as to effects due to the creation of trench sidewalls: a narrow range of trench depth values below which SA surface participation increases, followed by a regime in which contributions from the sidewalls saturate. For zero to intermediate trench depths, the ratio of SM surface participation to SA trench bottom surface participation is constant, dictated by the relative dielectric constants of the constituent materials. A comparison to experimentally measured quality factors extracted from trenched CPW resonators yields an upper bound on loss tangents comparable to that associated with oxide contamination on silicon substrates.

## Appendix A: Analytical approximation of $u'_i$ and $u_i$

The derivation of approximating the conformal transformation for cases of small trench depths follows the work of Gao for CPW designs with finite thickness metallization [27]. To produce an analytical approximation to the distance between $u'_1$ and $u_1$, we simplify the starting conformal map for $z$ in the neighborhood of $u'_1$:

$$\sqrt{\frac{(z+u_1)(z-u_1)(z+u_2)(z-u_2)}{(z+u'_1)(z-u'_1)(z+u'_2)(z-u'_2)}} \sim \sqrt{\frac{z-u_1}{z-u'_1}} \quad (19)$$

According to Fig. 1(b), the integration of (19) from $u'_1$ to $u_1$ corresponds to traversing the inner sidewall in Fig. 1(a) from 0 to $-i\,t$:

$$-it \sim \int_{u'_1}^{u_1} \sqrt{\frac{z-u_1}{z-u'_1}}\,dz = -(u_1-u'_1)\ln|i| = -i(u_1-u'_1)\frac{\pi}{2} \quad (20)$$

so that $u_1 \sim u'_1 + 2t/\pi$. Likewise, one can derive a similar approximation in the neighborhood of $u'_2$ to form $u'_2 \sim u_2 + 2t/\pi$.

This same procedure can be used to generate analytical approximations to the trench corner positions. Let us consider the region near the upper, left trench corner ($x \sim u'_1$). The conformal mapping dictated by (19) can be simplified to form:

$$w(z) \sim \int \sqrt{\left(\frac{z-u_1}{z-u'_1}\right)\left(1+\frac{d}{z+u'_1}\right)\left(1+\frac{d}{z-u'_2}\right)\left(1-\frac{d}{z+u'_2}\right)}\,dz \quad (21)$$

where $u'_1 \sim u_1 - d$ and $u'_2 \sim u_2 + d$. The binominal theorem can be applied:

$$w(z) \sim \int \sqrt{\frac{z-u_1}{z-u'_1}}\left[1+\frac{d}{2}\left\{\frac{1}{z+u'_1}+\frac{1}{z-u'_2}-\frac{1}{z+u'_2}\right\}\right]dz \quad (22)$$

Traveling in the $z$ plane from 0 to $u'_1$ corresponds to the horizontal distance $a$ in the complex $w$ plane:

$$a \sim \int_0^{u'_1} \sqrt{\frac{z-u_1}{z-u'_1}}\,dz + \frac{d}{2}\int_0^{u'_1}\left[\frac{1}{(z+u'_1)}+\frac{1}{(z-u'_2)}-\frac{1}{(z+u'_2)}\right]dz \quad (23)$$

Integrating the first term on the right-hand side of (23):

$$\int_0^{u'_1}\sqrt{\frac{z-u_1}{z-u'_1}}\,dz \sim u'_1 + \frac{d}{2}\left[1+\ln\left(\frac{4u'_1}{d}\right)\right]$$
$$\sim u'_1 + \frac{t}{\pi}\left[1+\ln\left(\frac{2\pi u'_1}{t}\right)\right] \quad (24)$$

Approximating the remainder of the right-hand side of (23):

$$\int_0^{u'_1}\frac{d}{2}\left\{\frac{1}{z+u'_1}+\frac{1}{z-u'_2}-\frac{1}{z+u'_2}\right\}dz = \frac{t}{\pi}\left[\ln(2)+\ln\left|\frac{u'_2-u'_1}{u'_2+u'_1}\right|\right] \quad (25)$$

Combining (24) and (25) and using the additional approximations for small trench depths ($u'_2 \sim b$ and $u'_1 \sim a$):

$$u'_1 \sim a - \frac{t}{\pi}\left[1+\ln(4\pi)+\ln\left(\frac{1-k}{1+k}\right)-\ln\left(\frac{t}{a}\right)\right] \quad (26)$$

Likewise, $u'_2$ can be approximated using a similar approach to form:

$$u'_2 \sim b + \frac{t}{\pi}\left[1+\ln(4\pi)+\ln\left(\frac{1-k}{1+k}\right)-\ln\left(\frac{t}{b}\right)\right] \quad (27)$$

For the coplanar capacitor design, as shown in the cross-sectional schematic Fig. 5, $u_i$ and $u'_i$ can be approximated using the same procedure to arrive at:

$$u'_1 \sim a + \frac{t}{\pi}\left[1+\ln(4\pi)+\ln\left(\frac{1-k}{1+k}\right)-\ln\left(\frac{t}{a}\right)\right] \quad (28)$$
$$u'_2 \sim b - \frac{t}{\pi}\left[1+\ln(4\pi)+\ln\left(\frac{1-k}{1+k}\right)-\ln\left(\frac{t}{b}\right)\right] \quad (29)$$

where $u_1 \sim u'_1 - 2t/\pi$ and $u_2 \sim u'_2 + 2t/\pi$.

## Appendix B: Analytical approximation of $\delta_1(x)$

Using the simplification of the conformal mapping from (4) for a region near $z = u'_1$, we can assess the path travelled in the complex $w$-plane corresponding to a vertical movement in the complex $z$-plane by a value of $\delta_1(x)$ for an arbitrary value of $x$:

$$\sqrt{\frac{(u_1^2-x^2)(u_2^2-x^2)}{(u_1'^2-x^2)(x+u'_1)}}\int_x^{x+i\delta_1}\frac{dz}{\sqrt{u'_1-x-iy}}$$
$$= 2\sqrt{\frac{(u_1^2-x^2)(u_2^2-x^2)}{(u_1'^2-x^2)(x+u'_1)}}\left[\sqrt{u'_1-x}-\sqrt{u'_1-x-i\delta_1(x)}\right]$$
$$= \alpha + i\,\delta_0 \quad (30)$$

where $\alpha$ and $\delta_0$ represent the horizontal and vertical components of movement in the $w$ plane. Equation (30) can be rearranged to form:

$$\sqrt{u'_1-x-i\delta_1(x)} = \sqrt{u'_1-x} - \frac{(\alpha+i\delta_0)}{2}\sqrt{\frac{(u_1'^2-x^2)(x+u'_1)}{(u_2^2-x^2)(u_1^2-x^2)}} \quad (31)$$

which, by squaring both sides and simplifying, produces the following relation for $\delta_1(x)$:

$$\delta_1(x) = \frac{[i(\alpha^2-\delta_0^2)-2\alpha\delta_0]}{4}\frac{(u_1'^2-x^2)(x+u'_1)}{(u_2^2-x^2)(u_1^2-x^2)}$$
$$-(i\alpha-\delta_0)\sqrt{\frac{(u_1'^2-x^2)(x+u'_1)}{(u_2^2-x^2)(u_1^2-x^2)}} \quad (32)$$

Since $\delta_1(x)$ must, by definition, be real, we can construct a quadratic equation for the unknown value $\alpha$ by setting the sum of the imaginary parts of (32) to be equal to zero:

$$\frac{(\alpha^2-\delta_0^2)}{4}\frac{(u_1'^2-x^2)(x+u'_1)}{(u_2^2-x^2)(u_1^2-x^2)}-\alpha\sqrt{\frac{(u_1'^2-x^2)(u_1'^2-x^2)}{(u_2^2-x^2)(u_1^2-x^2)}} = 0 \quad (33)$$

which possesses the solutions:

$$\alpha = 2\sqrt{\frac{(u_1^2-x^2)(u_2^2-x^2)(u_1'-x)}{(u_2'^2-x^2)(x+u_1')}}$$
$$\pm\sqrt{4\frac{(u_1^2-x^2)(u_2^2-x^2)(u_1'-x)}{(u_2'^2-x^2)(x+u_1')}+\delta_0^2} \qquad (34)$$

Taking the negative root of (34) for $\alpha$ and inserting this value back into (32), we arrive at an approximation for $\delta_1(x)$:

$$\delta_1(x)\sim\frac{\delta_0}{2}\left[\frac{(u_2^2-x^2)(u_1+x)}{(u_2^2-x^2)(u_1^2-x^2)}\right]\sqrt{\delta_0^2+4\frac{(u_2^2-x^2)(u_1^2-x^2)(u_1-x)}{(u_2^2-x^2)(u_1+x)}} \qquad (35)$$

Although this expression was generated under the assumption that $|x| \leq u'_1$, comparable formulas can be derived for the regimes in which $u'_1 < |x| < u_1$ and $|x| > u_1$. They all can be summarized using the following aggregate expression:

$$\delta_1(x)\sim\frac{\delta_0}{2}\left[\frac{(u_2^2-x^2)(u_1+x)}{(u_2^2-x^2)\mid u_1^2-x^2\mid}\right]\sqrt{\delta_0^2+4\frac{(u_2^2-x^2)\mid(u_1^2-x^2)(u_1-x)\mid}{(u_2^2-x^2)(u_1+x)}} \qquad (36)$$

A similar procedure can be employed to calculate the contamination layer thickness in the vicinity of $u'_2$, $\delta_2(x)$ to produce:

$$\delta_2(x)\sim\frac{\delta_0}{2}\left[\frac{(x^2-u_1^2)(x+u_2)}{(x^2-u_1^2)\mid x^2-u_2^2\mid}\right]\sqrt{\delta_0^2+4\frac{(x^2-u_1^2)\mid(x^2-u_2^2)(u_2-x)\mid}{(x^2-u_1^2)(x+u_2)}} \qquad (37)$$

Note that (36) and (37) are valid for both CPW and coplanar capacitor designs.